\documentstyle[prl,aps]{revtex} 
\begin{document}
\title{Weak magnetism for antineutrinos in supernovae}
\author{C.~J.~Horowitz
\footnote{e-mail:  charlie@iucf.indiana.edu} 
}
\address{Nuclear Theory Center and Dept. of Physics, Indiana
University, Bloomington, IN 47405}
\date{\today} 
\maketitle 
\begin{abstract}
Weak magnetism increases antineutrino mean free paths in core collapse 
supernovae.   The parity violating interference between axial and vector 
currents makes antineutrino-nucleon cross sections smaller then those for 
neutrinos.  We calculate simple, exact correction factors to include recoil 
and weak magnetism in supernova simulations.   Weak magnetism may 
significantly increase the neutrino energy flux.  We calculate, in a 
diffusion approximation, an increase of order 15\%
in the total energy flux for temperatures near 10 MeV. This should raise 
the neutrino luminosity. 
Weak magnetism also changes the emitted spectrum of 
$ \bar\nu_x$ (with $x=\mu$ or $\tau$) and $\bar\nu_e$. We estimate that 
$\bar\nu_x$ will be emitted about 7\% hotter than $\nu_x$ because 
$\bar\nu_x$ have longer mean free paths.  Likewise weak magnetism may 
increase the $\bar\nu_e$ temperature by of order 10\%.  This increase in 
temperature coupled with the increase in neutrino luminosity should increase
the heating in the low density region outside of the neutrino sphere.
This, in turn, could be important for the success of an explosion. 
It is important to check our results with a full simulation that includes 
Boltzmann neutrino transport and weak magnetism corrections.  

\end{abstract}


\section{Introduction}

Core collapse supernovae are dominated by neutrinos, therfore many supernova properties may depend on the nature of neutrino-nucleon interactions.  This provides an opportunity to study characteristic symmetries and features of the Standard Model weak interactions.  Supernovae may provide macroscopic manifestations of charge conjugation \cite{hor98,hor00} or parity violation \cite{vil79,hor98b,arr99}.

In the laboratory at high energies, antineutrino-nucleon cross sections
are systematically smaller than neutrino-nucleon cross sections. 
This is related to charge conjugation and parity violation in the
Standard Model.  Nevertheless, most core collapse supernova simulations
still use the {\it same} lowest order cross section $d\sigma_0/d\Omega$ for both neutrinos and antineutrinos,
\begin{equation}
{d\sigma_0 \over d\Omega} = {G^2k^2\over 4\pi^2} [ c_v^2(1+{\rm cos}\theta)+c_a^2(3-{\rm cos}\theta)].
\label{1}
\end{equation}
Here $k$ is the (incoming) neutrino energy, $\theta$ the scattering angle and 
the vector $c_v$ and axial vector $c_a$ coupling constants are listed
in table I.

In this paper, we discuss free space corrections to Eq. (\ref{1}) from 
nucleon recoil, weak magnetism, strange quarks and single nucleon form factors. 
These corrections may be important for supernova simulations because they are present 
at all densities. In contrast, the density dependent nucleon correlations discussed by \cite{red99} 
and \cite{bur98} are important only at very high densities, well inside the neutrino sphere.

Electron antineutrinos were detected from Supernova 1987A via their capture on protons.  
Weak magnetism reduces this cross section and impacts the deduced neutron star binding
energy and antineutrino temperature.  Weak magnetism also changes the rate 
of antineutrino capture in the neutrino driven wind above a protoneutron star. 
This may change the electron fraction $Y_e$ (ratio of electrons 
or protons to baryons) and nucleosynthesis yields in the wind \cite{hor99,hor01}.

However, to confirm this change in $Y_e$, one should perform supernova 
simulations including weak magnetism because this may change the neutrino 
luminosities and emitted spectra which will also impact $Y_e$.  In this paper,
we present simple formulae so that these corrections can be incorporated in simulations. 
We also give estimates of the change in emitted electron antineutrino spectrum
and the differences expected between the spectrum of mu or tau antineutrinos and neutrinos.

The reduction in opacities from recoil and weak magnetism should increase 
luminosities of both neutrinos and antineutrinos as more energy is transported
to the neutrino sphere.  This could increase heating behind the shock 
and more then compensate for the smaller antineutrino absorption cross section. 
Weak magnetism should also reduce the cooling from positron capture on 
neutrons further increasing the net heating behind the shock. 
These changes in heating could make a simulation more likely to explode.

Finally, weak magnetism allows mu and tau antineutrinos to escape the 
star faster than mu and tau neutrinos.  This should lead to a large muon number 
(number of mu neutrinos minus antineutrinos) or tau number of up to $10^{54}$ 
for the hot protoneutron star \cite{hor98}.  This is accompanied 
by a nonzero chemical potential for mu and tau neutrinos.

Accurate treatments of neutrino transport are now feasible based on the Boltzmann equation \cite{ram00}.  Therefore, it is important to systematically improve neutrino opacities.  In section II, we catalog a number of possible opacity corrections.

In section III, we present exact neutrino-nucleon cross sections for both charged
and neutral currents.  These are accurate to all orders in $k/M$ where $k$ is 
the neutrino energy and $M$ the nucleon mass.  We also present 
corrections from single nucleon form factors and strange quark contributions 
in the nucleon.  In section IV we discuss recoil and weak magnetism corrections to 
mu and tau neutrino, energy and lepton number fluxes and spectra.  
Section V discusses corrections for electron antineutrinos.  We also discuss possible changes in electron fraction $Y_e$ and heating rates.  Finally, we conclude in section VI that these weak magnetism and recoil corrections should be incorporated in future supernova simulations.

\section{Corrections to Neutrino Opacities}

Recently, accurate numerical algorithms have been developed 
to solve the Boltzmann equation for neutrino transport \cite{ram00}.  
The high accuracy of these simulations now warrants
systematic improvements in the neutrino opacities.  
Therefore, we discuss a number of opacity corrections.

The corrections listed in table II can be divided into
two main groups. Numbers 1 through 5 are classified as model 
independent because they can be calculated exactly 
and are independent of the model used to describe dense matter. 
In contrast, corrections 6 through 13 are model dependent. 
Not only do these corrections depend on the model,
but it is important that these corrections be  
consistent with the equation of state.

Corrections 1, phase space, and 2, matrix element are perhaps
the simplest.  These are the only ones that are present in 
the limit of very low densities.  They represent more accurate
calculations of free space neutrino-nucleon scattering.  By 
phase space we mean corrections from the $Q$ value of the 
reaction such as the neutron-proton mass difference or 
because the outgoing neutrino momentum is slightly different
from the incoming momentum.

In this paper we focus on number 2, corrections to the matrix
element. Recoil corrections, 2a, arise because the nucleon 
is not infinity heavy and recoils slightly from
the neutrino.  Weak magnetism, 2b, arise from the parity
violating interference between the weak magnetic moment of a 
nucleon and its axial current.  Weak magnetism is important because it has opposite sign
for neutrinos and antineutrinos. Thus weak magnetism
increases the opacity for neutrinos and decreases the opacity
for antineutrinos.

The first effects of the dense medium are 3, Pauli blocking 
of some outgoing nucleon or electron states, and 4, the Fermi
or thermal motion of the initial nucleon or electron. 
These can be calculated exactly. The final model independent correction in table II is 5, 
coulomb corrections.  If heavy nuclei are present, the 
opacity may be dominated by coherent elastic scattering 
from nuclei. However, these nuclei have strong coulomb 
interactions and form a strongly correlated classical liquid.
The correlations between ions can greatly reduce the opacity
for low energy neutrinos that have wavelengths comparable 
to the distance between ions.  This correction is model independent
because the coulomb interaction is known.  For example, 
Ref. \cite{hor97}  calculated the exact static structure factor
of the ions with a simple Monte Carlo simulation.  In principle,
coulomb correlations are also present for neutrino electron 
scattering \cite{hor91}.

Corrections 6--13 depend on the model used for the 
strong interactions between nucleons.  The mean field
approximation, 6, is perhaps the simplest way to treat the 
interactions.  Nonrelativistic mean field models 
incorporate mean field effects in the density of states 
with an effective mass.  Relativistic mean field models have both
Lorentz scalar and vector mean fields and the scalar field
changes the effective Dirac mass.

It is important that the opacities be consistent with the model
used for the equation of state (EOS). This can be
achieved by using linear response theory \cite{fet71} to
calculate the response of the medium to a neutrino probe.  
The random phase approximation (RPA) or sum of ring 
diagrams provides the linear response of a mean field theory
ground state.  Therefore, if one uses a mean field EOS and
calculates the opacities in RPA, where the RPA effective
interaction is the same as that used for the EOS, then 
the opacities and EOS will be consistent.  For example, 
the relativistic RPA opacities of Ref. \cite{red99} are
consistent with their mean field EOS.

We note that the nonrelativistic calculations of Burrows
and Sawyer \cite{bur98} include corrections 1, 3--7 
however they do not contain any matrix element correction, 2.
Likewise the relativistic RPA calculations of \cite{red99}
contain corrections 1, 3--7 and 2a recoil but do not contain 
most of 2b weak magnetism.  In the present paper we discuss
weak magnetism corrections. In later work we will 
incorporate these into full nonrelativistic and relativistic 
RPA calculations.

There are many $NN$ correlations beyond a simple RPA 
approximation, 8, for example those responsible
for $NN \leftrightarrow NN\nu\bar\nu $ \cite{han98}.
In addition there are meson exchange current corrections, 9.
The Stony Brook group has done some work calculating 
corrections from hyperons or other strange hadron 
components, 10, \cite{red98}.  More work could be done calculating 
opacities for meson condensates or other exotic phases of dense
matter, 11.

If superfluid or superconducting phases are present  there
will be pairing corrections to the opacity, 12.  These can 
be important for the late time cooling of a neutron star \cite{flo76}
or perhaps if there is a color superconducting phase \cite{car00}.
Finally there could be important corrections to opacities
in nonuniform matter, 13.  For example, in the low density
pasta phases present in a neutron star inner crust or for a 
nonuniform meson condensate\cite{red00}.

All of the above assumed magnetic fields do not play
important roles.  If magnetic fields are important, 
they could substantially complicate the opacity, 14,
see for example \cite{har01}.
Magnetic fields might be important in jets or to break
the spatial symmetry to explain neutron star kicks \cite{vil79,hor98b,arr99}.

To conclude this section, now that accurate neutrino transport
is possible, it is important to systematically improve  the 
neutrino opacities. One should include weak magnetism
and other free space corrections along with the many 
density dependent effects.  
Furthermore, these free space corrections 
can be included exactly and without any model dependence.

\section{Neutrino nucleon scattering}

In this section we discuss recoil, weak magnetism,
nucleon form factor, and strange quark corrections to the zeroth order cross section, Eq. 
(\ref{1}).  Then we present differential, transport and total cross sections that are exact 
to all orders in the neutrino energy over the nucleon mass $k/M$.  
\medskip
\subsection{Phase space}

Perhaps the simplest correction comes from the phase space of the scattered 
lepton because its energy $k'$ is less than
the incident energy $k$.  If we use full relativistic 
kinematics for the nucleon and we assume it is originally 
at rest, then, 
\begin{equation}
k' = k/[1+e(1-x)] \equiv  \eta k , 
\label{2}
\end{equation}
\begin{equation}
\eta = [1+e (1-x)]^{-1},
\label{3}
\end{equation}
with $x= cos \theta$.

We expect corrections to depend on the small parameter, 
\begin{equation}
e \equiv k/M .
\label{4}
\end{equation}
To the first order in $e$, 
\begin{equation} 
k' \approx k [ 1-e (1-x)] . \label{5}
\end{equation}
If one uses the phase space with $k'$ but still uses 
the matrix element for $M \rightarrow \infty $ then Eq.
(\ref{1}) becomes,
\begin{equation}
{d\sigma_{ps}\over d\Omega} = 
{G^2k'^2\over 4\pi^2} \left[ c_v^2 (1 + x) +
c_a^2 (3-x)\right], 
\label{6}
\end{equation}
\begin{equation}
= {d\sigma_0 \over d\Omega} \eta^2 ,
\label{7}
\end{equation}
\begin{equation}
\approx {d\sigma_0 \over d\Omega} \left[1-2e(1-x)\right] .
\label{8}\\
\end{equation}
We note that Burrows and Sawyer's work \cite{bur98} in the limit
of low densities reduces to $d\sigma_{ps}\over d\Omega$, 
see Appendix I. 
\medskip

\subsection{Recoil}

If one evaluates the matrix element to first order in e and
includes the phase space corrections one has, 
\begin{equation}
{d\sigma_R\over d\Omega} \approx {d\sigma_0 \over d\Omega}
\left[1-3e (1-x)\right] . 
\label{9}
\end{equation}
We note that the full recoil correction in Eq. (\ref{9}) is 50\%
larger than that for $d\sigma_{ps}\over d\Omega$ in Eq. (\ref{8}).
Equation (9) still neglects the parity violating interference between 
axial and vector currents.  This interference is dominated by weak magnetism.
\medskip
\subsection{Weak Magnetism}

The full weak vector current $J_\mu$ has Dirac $(c_v)$ and Pauli 
or tensor $(F_2)$ contributions,
\begin{equation}
J_\mu=c_v \gamma_\mu + F_2 {i\sigma_{\mu\nu}q^\nu
\over 2M}\label{10}
\end{equation}
where $q_\mu= k_\mu - k'_\mu $ 
is the momentum transfered to the nucleon.  The weak current 
$J_\mu$ is related by CVC (conserved vector current) to the 
electromagnetic current. Therefore the large anomalous magnetic moments of 
the proton and neutron give rise to a large $F_2$. The couplings
$c_v$ and $F_2$ are collected in table I. 

The cross section to the first order in $e$ is, 
\begin{equation}
{d\sigma \over d\Omega} \approx {d\sigma_0 \over d\Omega} \left[
1+ \left( \pm {4c_a (c_v + F_2) \over c^2_v (1+x) + c^2_a(3- x)}
-3\right) e (1-x)\right],
\label{11}
\end{equation}
with plus sign for $\nu$ and minus sign for $\bar\nu$. Weak
magnetism increases $\nu$ and decreases $\bar\nu$, while
recoil decreases both $\nu$ and $\bar\nu$ cross sections.
Therefore weak magnetism and recoil corrections approximately 
cancel for $\nu$ but add for $\bar\nu$.  Note, Burrows et al. 
\cite{burrows00}, see also for example \cite{vog99}, include 
weak magnetism for the charged currents but neglect it for 
neutral currents.  Equation (11) is a good approximation for 
$\nu$. However it can become negative for large $\bar\nu$ energies. 
To cure this problem we include corrections to all orders in $e$. 
\medskip

\subsection{Exact cross section}

The exact cross section can be calculated, see also \cite{ahr87},
\begin{eqnarray}
{d\sigma\over d\Omega} = {G^2k^2\over 4 \pi^2} \eta^2
\Biggl\{ 
c_v^2 \left[ 1+ \eta^2 - \eta (1-x) \right] + c_a^2 
\left[ 1+\eta^2 + \eta (1-x) \right] 
\pm 2c_a (c_v+F_2) (1-\eta^2)\nonumber \\
\left. + \eta^2 {e^2\over 2} (1-x) \left[F^2_2 (3-x) + 4c_v F_2 (1-x)
\right] \right\} . 
\label{12}
\end{eqnarray}

In general the $F^2_2$ and $c_vF_2$ terms are small because
they only enter at order $e^2$. However, their inclusion is
important to ensure a positive cross section at high energies.  Note, Schinde considers the $c_ac_v$ term but neglected all $F_2$ contributions \cite{sch90}.

\medskip
\subsection{Nucleon Form Factors}

The finite size of the nucleon can be included in Eq. (\ref{12}) by using
appropriate form factors, 
\begin{eqnarray}
F_2 \rightarrow F_2 (Q^2),\nonumber \\
c_v \rightarrow c_v (Q^2), \nonumber \\
c_a \rightarrow c_a (Q^2),
\label{13}
\end{eqnarray}
that are functions of the four momentum transfer squared,
\begin{equation}
Q^2 =-q^2_\mu = 2k^2\eta (1-x) .
\label{14}
\end{equation}
 For example, for $\nu_e n\rightarrow e^-p$, 
\begin{equation}
c_a (Q^2) \approx g_a / (1+Q^2/M^2_A)^2 ,
\label{15}
\end{equation}
with $g_a \approx$ 1.26 and the axial mass $M_A \approx$ 1 GeV.
Form factors only enter at order $e^2$. They reduce
the cross section by about 10\% at $k\approx 100 MeV$,
and even less at lower energies. However, 
for completeness we include simple parameterizations 
of the form factors in Appendix II. Because the corrections are so 
small, we will ignore the form factors in the rest of
this paper.  

\medskip
\subsection{Total charged current cross section}

The opacity for $\nu_e$ from charge current reactions depends
on the total cross section for $\nu_e n \rightarrow e^- p$,
\begin{equation}
\sigma = \int {d\sigma\over d\Omega} d\Omega .
\label{16}
\end{equation}
The zeroth order result is, 
\begin{equation}
\sigma_0 = {G^2k^2\over \pi} (c_v^2 + 3 c_a^2) ,
\label{17}
\end{equation}
while the correction factor from just phase space is,
\begin{equation}
\sigma_{ps} =\int {d\sigma_{ps}
\over \sigma\Omega} d\Omega =\sigma_0 R_{ps}(k) , \label{18}
\end{equation}
\begin{equation}
R_{ps}(k) =\left\{ c_v^2 \left[ {1\over e} - {1\over 2e^2}
ln(1+2e)\right]+c_a^2 
\left[ {2e-1 \over e(1+2e)} + {1\over 2e^2}
ln (1+2e)\right]\right\} / (c_v^2 +3c_a^2) , \label{19}
\end{equation}
\begin{equation}
R_{ps} \approx  1 - { 4\over 3} \left( {c_v^2 + 5c_a^2 
\over c_v^2 + 3c_a^2} \right) e     + O(e^2) . \label{20}
\end{equation}
The exact result is, 
\begin{equation}
\sigma = \int {d\sigma \over d\Omega} d\Omega = \sigma_0
R(k) , \label{21}
\end{equation}
\begin{eqnarray}
R(k) = \left\{ c_v^2 \left( 1 + 4e + {16 \over 3} e^2 \right)
+ 3c_a^2 \left( 1 + {4\over 3}e \right)^2 \pm 
4\left( c_v + F_2 \right) c_a e \left(1+ {4 \over 3}e \right)
\right. \nonumber \\
\left. + {8 \over 3} c_v F_2 e^2 + {5 \over 3} e^2 \left( 1+ {2 \over 5} e
\right) F^2_2  \right\} / \left[ \left(c_v^2 + 3 c_a^2 \right)
 (1+ 2e)^3 \right] ,
\label{22}
\end{eqnarray}
\begin{equation}
R(k) \approx 1-2 \left( {c_v^2 + 5c_a^2 \bar + 2 \left(c_v 
+ F_2 \right) c_a \over c_v^2 + 3c_a^2 } \right) e 
+ O (e^2). \label{23}
\end{equation}
For $\bar\nu_e$ using table I this is,
\begin{equation}
R_{\bar\nu_e} \approx 1- 7.22e \label{24}
\end{equation}
while for $\nu_e$ we have, 
\begin{equation}
R_{\nu_e} \approx 1 + 1.01 e \label{25}
\end{equation}
We note the main effect of weak magnetism and recoil is to reduce
the $\bar \nu _e$ opacity by the large amount in Eq. (\ref{24}).
For a typical neutrino energy $k\sim <E>\sim$ 20 MeV this is a 
15\% reduction. 

\begin{figure}
\vbox to 5in{\vss\hbox to 8in{\hss {\includegraphics{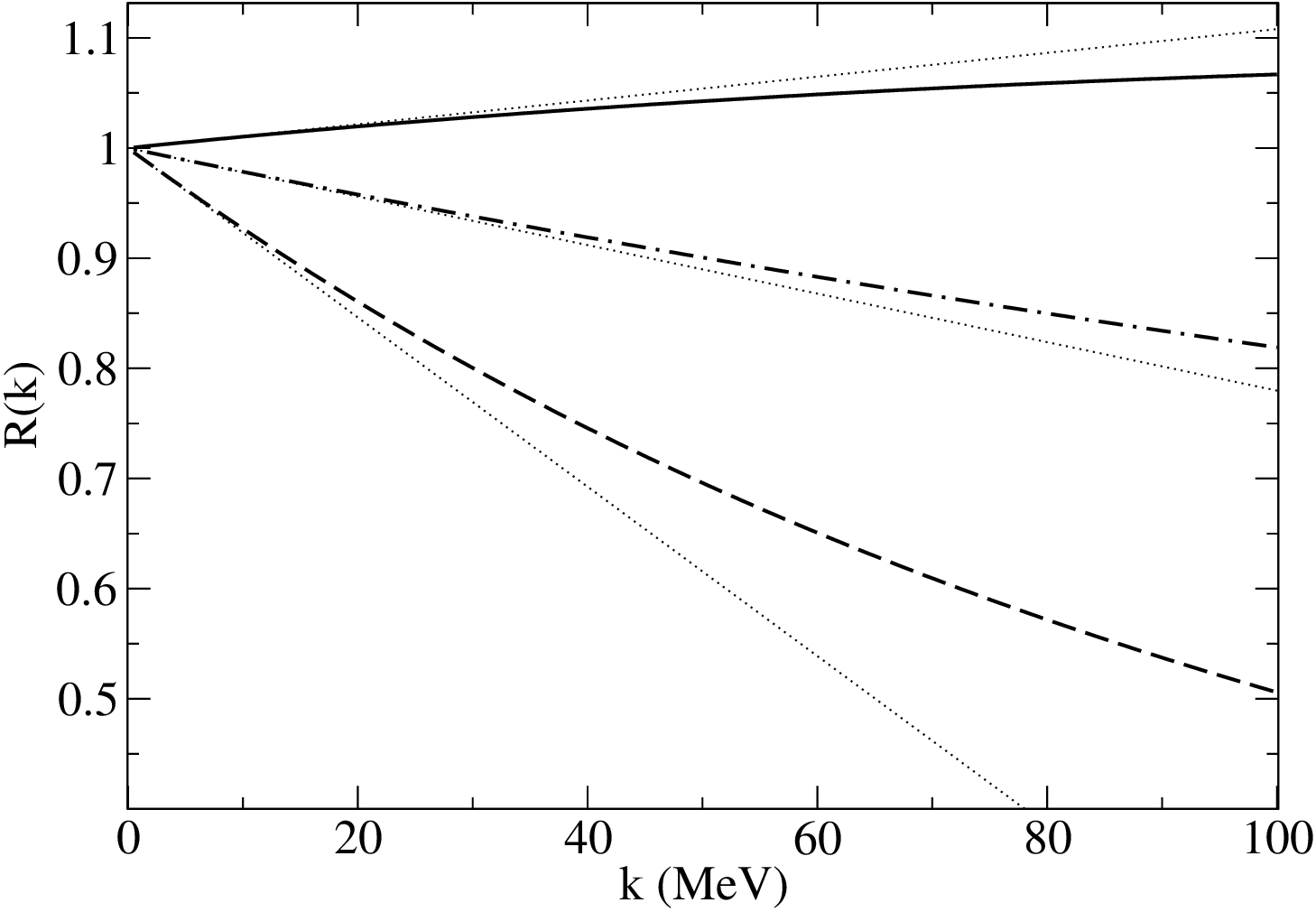}}\hss}}
\caption{Factor $R(k)$ that corrects the total charged current
cross section for weak magnetism and recoil, Eq. (\ref{22}), versus neutrino
energy $k$.  Solid line is for neutrinos, dashed line is for 
antineutrinos and the dot-dashed line includes only phase
space corrections, Eq. (\ref{19}).  Also shown as dotted lines
are the corresponding lowest order correction factors, Eqs. (\ref{20}) and (\ref{23}). }
\label{fig1}
\end{figure}

The full correction factor $R(k)$, is plotted in Fig. \ref{fig1}
along with the lowest order forms in Eqs. (\ref{24},\ref{25}). We see
that the lowest order is fine for neutrinos but fails for antineutrinos
above about 50 MeV.

\medskip
\subsection{Neutral current transport cross section}

For mu and tau neutrinos the scattering opacity depends on the transport
cross section,
\begin{equation}
\sigma^t = \int d\Omega {d\sigma \over d\Omega} \left(1-x\right) .
\label{26}
\end{equation}
Note, strictly speaking this is true only in the limit
$k^\prime =k$ nevertheless Eq. (\ref{26}) provides a good estimate when
$k^\prime\approx k$.

\begin{figure}
\vbox to 5in{\vss\hbox to 8in{\hss {\includegraphics{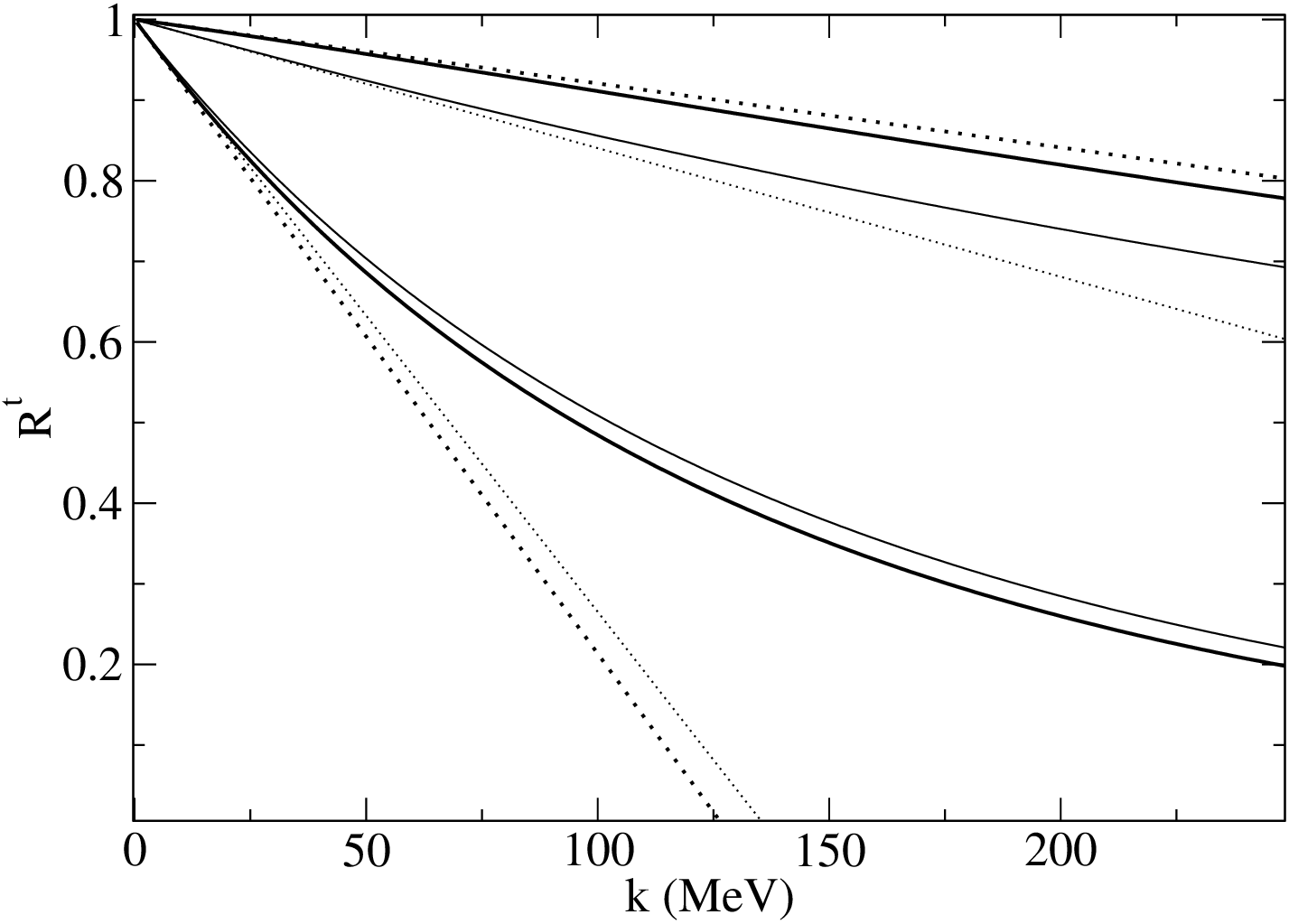}}\hss}}
\caption{Factor $R^t(k)$, that 
corrects transport cross sections for
weak magnetism and recoil, Eq. (\ref{32}), versus neutrino energy $k$.
The upper thick solid line is for $\nu n\rightarrow \nu n$
while the upper thin solid line is for $\nu p \rightarrow \nu p$.
The corresponding lower solid curves are for antineutrino
scattering while the dotted curves give lowest order results,
Eq. (\ref{33})}
\label{fig2}
\end{figure}

The zeroth order transport cross section is,
\begin{equation}
\sigma_0^t = {G^2k^2\over \pi} {2\over 3} \left(c_v^2 + 5 c_a^2\right) .
\label{27}
\end{equation}
The transport cross section including only phase space corrections is,
\begin{equation}
\sigma_{ps}^t = \sigma^t_0 R^t_{ps} (k), \label{28}
\end{equation}
\begin{eqnarray}
R^t_{ps}(k) = \left\{ c_v^2 \left[ {\left( 1+e \right) \over e^3}
ln \left( 1+2e \right) - {2\over e^2} \right] + c^2_a
\left[ {2e +\left(2e^2 -e-1\right) ln \left(1 + 2e\right)
\over e^3 \left(1+2e\right)} \right] \right\} \nonumber \\
/ \left[ {2\over 3} \left( c_v^2 + 5c_a^2 \right)\right] ,
\label{29}
\end{eqnarray}
\begin{equation}
R^t_{ps}\approx 1- \left( {2 c_v^2 + 14 c_a^2 \over
c_v^2 + 5c_a^2}\right) e+O(e^2) . \label{30}
\end{equation}
Finally the exact transport cross section is 
\begin{equation}
\sigma^t = \sigma^t_0 \ R^t(k), \label{31}
\end{equation}
\begin{eqnarray}
R^t(k) =\left\{ c_v^2 \left[ {e-1\over 2e^3} ln (1+2e) + 
{3+ 12e + 9e^2 - 10e^3 \over 3e^2 \left( 1+2e \right)^3}
\right] \right. \nonumber \\ 
+c_a^2 \left[ {1+e\over 2e^3} ln (1+2e) - 
{10e^3 +27e^2 +18e +3 \over 3e^2 (1+2e)^3}\right] \nonumber \\
\pm  \left(c_v +F_2\right) c_a \left[ {1\over e^2} ln \left(
1+2e\right) - 
{2+10e+ {28\over 3}e^2 \over e (1+2e)^3} \right] \nonumber \\
+ c_v F_2 \left[{1\over e^2}ln (1+2e) -{2\over 3} 
\left( {3+15e +22e^2\over e (1+2e)^3}\right)\right]\nonumber \\
\left. + F^2_2 \left[{1\over 4e^2} ln (1+2e) + 
{8e^3 -22e^2 - 15e-3 \over 6e (1+2e)^3} \right] \right\} / \nonumber \\
\left[ {2\over 3} \left(c^2_v +5c_a^2 \right) \right] , 
\label{32}
\end{eqnarray}
\begin{equation}
R^t \approx 1+\left(
{-(3c_v^2 + 21c_a^2) \pm 8 \left(c_v+F_2\right) c_a \over
c_v^2 +5c_a^2} \right) e+O(e^2). \label{33}
\end{equation}
The full correction, Eq. (\ref{32}), is plotted in Fig. (\ref{fig2}).  Corrections for
$\nu-n$ are very close to those for $\nu-p$ elastic scattering.   
We also plot the lowest order result, Eq. (\ref{33}).  This fails for antineutrinos at high energies. 
\medskip
\subsection{Strange Quark Contributions}
Neutrino nucleon elastic scattering is sensitive to possible strange quark contributions in the nucleon.  Strange quarks do not contribute to $c_v(Q^2=0)$ because the nucleon has no net strangeness.  However both $F_2(Q^2=0)$ and $c_a(Q^2=0)$ can be modified.  Parity violating electron scattering constrains $F_2$ \cite{sample}.  In this section we concentrate on $c_a$ because $F_2$ only contributes at order $k/M$.

Strange quark contributions are assumed to be isoscalar while the dominant contribution to $c_a$ is isovector.  Therefore, strange quarks are expected to increase the cross section for $\nu-p$ scattering and decrease that for $\nu-n$.  We write,
\begin{equation}
c_a(Q^2=0)= {1\over 2}(\pm g_a - g_a^s),
\end{equation}
with $g_a=1.26$ and the plus sign for $\nu-p$ and the minus sign for $\nu-n$ scattering.  The strange quark contribution $g_a^s$ is expected to be negative or zero.  The best limit comes from a Brookhaven $\nu-p$ scattering experiment \cite{ahr87}.  We assume,
\begin{equation}
g_a^s\approx -0.1 \pm 0.1.
\end{equation}
A better measurement of $g_a^s$ from a laboratory experiment would be very useful.

We make a simple estimate of strange quark effects starting from the zeroth order transport cross section in Eq. (\ref{27}).  We consider the neutral current opacity for a mixture of neutrons and protons of electron fraction $Y_e$.  For simplicity we set $c_v=0$ for $\nu-p$ scattering.  The ratio of the opacity with $g_a^s$ to that with $g_a^s=0$ is,
\begin{equation}
S(g_a^s)={5[g_a^2+(g_a^s)^2] + 10 g_a g_a^s(1-2Y_e)+1-Y_e\over
5g_a^2+1-Y_e}.
\end{equation}
For $g_a^s=-0.2$, $S$ represents a 21\% reduction in the opacity at $Y_e=0.1$ or a 15\% reduction at $Y_e=0.2$.  Thus strange quarks could reduce opacities by 10 to 20\%.  Note, strange quarks do not contributed to charged current interactions.

\medskip
\section{Muon and Tau Neutrinos}

In this section, we use the weak magnetism and recoil corrections for 
transport cross sections, Eq. (\ref{32}) to discuss 
muon and tau neutrino properties in supernovae. 
First, we examine muon and tau lepton number currents. Because neutrinos 
now have shorter mean free paths than antineutrinos, there will be net 
$\nu_\mu$ and $\nu_\tau$ densities and nonzero $\nu_\mu$ and $\nu_\tau$ chemical potentials.
We then calculate the increase in the $\nu_\mu$ 
and $\nu_\tau$ energy flux because of weak magnetism and recoil.
This should increase the neutrino luminosities.  Finally we
examine corrections to the emitted spectra of $\nu_\mu$, $\nu_\tau$
and $\bar\nu_\mu$  $\bar\nu_\tau$. We show that the 
$\bar\nu_\mu, \bar\nu_\tau$ are expected to be hotter than 
$\nu_\mu , \nu_\tau$ because of weak magnetism. 

For simplicity, we neglect charged muons. These could play some role 
at very high densities   and or temperatures.  Without muons,
our results are identical for muon and tau neutrinos.  Therefore,
we write $\nu_x$ in this section where $x$ could be $\mu$ or $\tau$.
\medskip
\subsection{Lepton number currents}

We start by considering high densities, well inside the neutrino sphere, 
where a simple diffusion approximation is valid.
Earlier work \cite{hor98,hor00} calculated the lepton number current
to first order in $e=k/m$.  We extend this work to all orders.

The transport mean free path for a $\nu_x$ of energy $k$ is,
\begin{equation}
\lambda (k) = \lambda_0 {k_0^2\over k^2} {1\over {R^t_\nu(k)}} ,
\label{34}
\end{equation} 
with the zeroth order mean free path $\lambda_0$ evaluated at an 
arbitrary reference energy $k_0$, 
\begin{equation}
\lambda_0 =\left[\sigma_0^t (k_0) \rho_n \right]^{-1},
\label{35}
\end{equation}
and $\rho_n$ is the neutron number density.
Note, for simplicity we assume pure neutron matter,
$Y_e=0$. Because the correction factor $R^t_\nu$, Eq.~(\ref{32}), is so similar for $\nu -p$ and $\nu - n$
scattering our results are not expected to change much for $Y_e > 0$.

The lepton number current for $\nu_x $ in a simple diffusion 
approximation is, 
\begin{equation}
\vec J_{\nu_x} =-\int {d^3k \over (2\pi)^3}
{\lambda (k) \over 3} \vec\nabla 
{1 \over  1 + {\rm exp} \left[ (k-\mu_{\nu_x}) / T \right]}.
\label{36}
\end{equation}
Here $\mu_{\nu_x}$ is the neutrino chemical potential. 
For $\bar\nu_x$ we have, 
\begin{equation}
\bar\lambda (k) = \lambda_0 {k_0^2\over k^2}
{1 \over R^t_{\bar\nu} (k)} , 
\label{37}
\end{equation}
This is now longer then $\lambda (k)$ because of the different
weak magnetism correction factors $R^t_{\bar\nu} < R^t_\nu$.
The $\bar\nu_x$ lepton number current has the same form 
as Eq. (\ref{36}) with $\mu_{\nu_x}$ replaced by $-\mu_{\nu_x}$,
\begin{equation}
\vec J_{\bar\nu_x} = - \int { d^3 k\over (2\pi )^3 }
{ \bar\lambda (k) \over 3 } \vec\nabla 
{1\over 1 + {\rm exp} \left[ (k + \mu_{\nu_{x}} )/T \right]} . 
\label{38}
\end{equation}
The total lepton number current is,

\begin{eqnarray}
\vec J = \vec J_{\nu_{x}} - \vec J_{\bar\nu_{x}} = -
{\lambda_0 k^2_0 \vec\nabla \over 6\pi^2} 
\left[ \int^\infty_0 {dk \over R^t_\nu} 
{ 1 \over {1 + {\rm exp} \left[ ( k- \mu_{\nu_{x}} ) /T \right]}} 
\right. \nonumber \\
\left. - \int^\infty_0 {dk \over R^t_{\bar\nu} } 
{1 \over 1 + {\rm exp} \left[ (k+\mu_{\nu_{x}}) /T \right] } \right]. 
\label{39}
\end{eqnarray}
If $\vec J \ne 0$ the lepton number of the star 
(number of $\nu_x$ - number of $\bar\nu_x$) will rapidly change. 
This buildup in lepton  number gives rise to a nonzero 
chemical potential $\mu_{\nu_x}$.

\begin{figure}
\vbox to 5in{\vss\hbox to 8in{\hss {\includegraphics{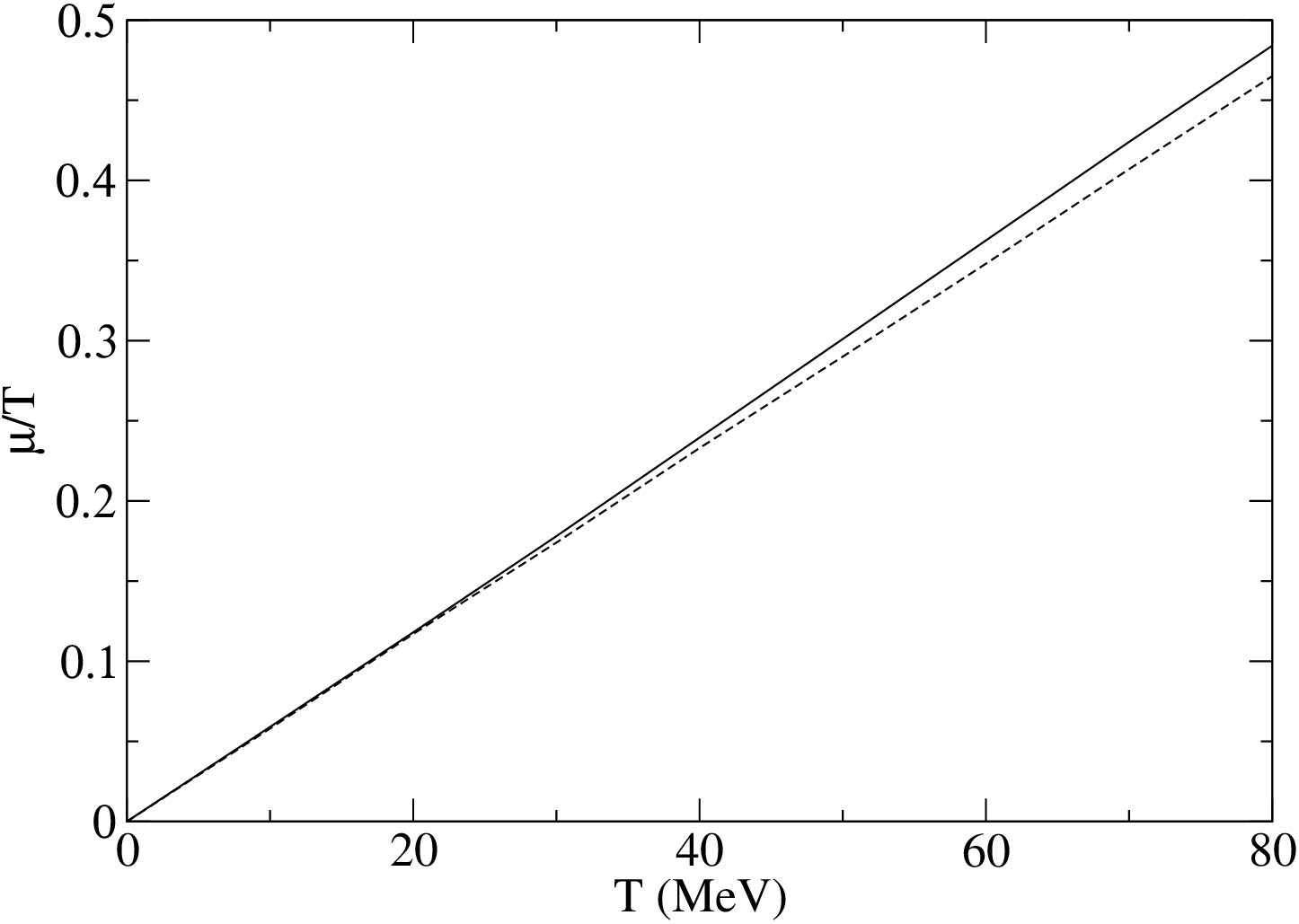}}\hss}}
\caption{Muon or Tau neutrino chemical potential over
temperature $\mu_{\nu_x} /T$ versus $T$ for matter in 
steady state equilibrium.  The solid line is the full
result from the solution to Eq. (\ref{40}) while the 
dashed line is correct to lowest order in $k/M$ and 
$\mu / T$, Eq. (\ref{41})  }
\label{fig3}
\end{figure}

Reference \cite{hor98} argued that the system rapidly reaches steady state
 equilibrium where $\vec J = \vec J_{\nu_x} - \vec J_{\bar\nu_x} =0$.
We numerically solve for the chemical potential $\mu_{\nu_x}$ so that,
\begin{equation}
\int^\infty_0 {dk\over R^t_\nu(k)} {1 \over 1+e^{(k-\mu_{\nu_x})/T} }
= \int^\infty_0 {dk\over R^t_{\bar\nu}(k)} {1 \over 1+e^{(k+\mu_{\nu_x})/T} }
\label{40}
\end{equation}
This is shown in Fig. \ref{fig3} as a function of $T$. This chemical
potential insures $J=0$.

If we expand Eq. (\ref{40}) to first order in $\mu_{\nu_x}/T $
and use $R^t_\nu (k)$  and $R^t_{\bar\nu} (k)$
expanded to the first order in $k/M$ one reproduces the lowest
order chemical pot $\mu_0$ of ref. \cite{hor98}, 
\begin{equation}
\mu_{\nu_x} \approx \mu_0 = {\delta _c \pi^2\over 6}
{T^2\over M}, \label{41}
\end{equation}
with $\delta_c = 8 (c_v +F^2) c_a / (c_v^2 + 5 c^2_a) \approx 3.30$
for $\nu - n$ scattering.  The lowest order $\mu_0$ is seen in Fig.
\ref{fig3} to be an excellent approximation to the exact 
result even at $T=80$ MeV.  Note, this nonzero $\mu_{\nu_x}$
can lead to muon number (number of $\nu_\mu$ minus number of
$\bar\nu_\mu$) or tau number for the protoneutron star as 
large as $10^{54}$ \cite{hor98}.
\medskip

\subsection{Energy Flux}

\begin{figure}
\vbox to 5in{\vss\hbox to 8in{\hss {\includegraphics{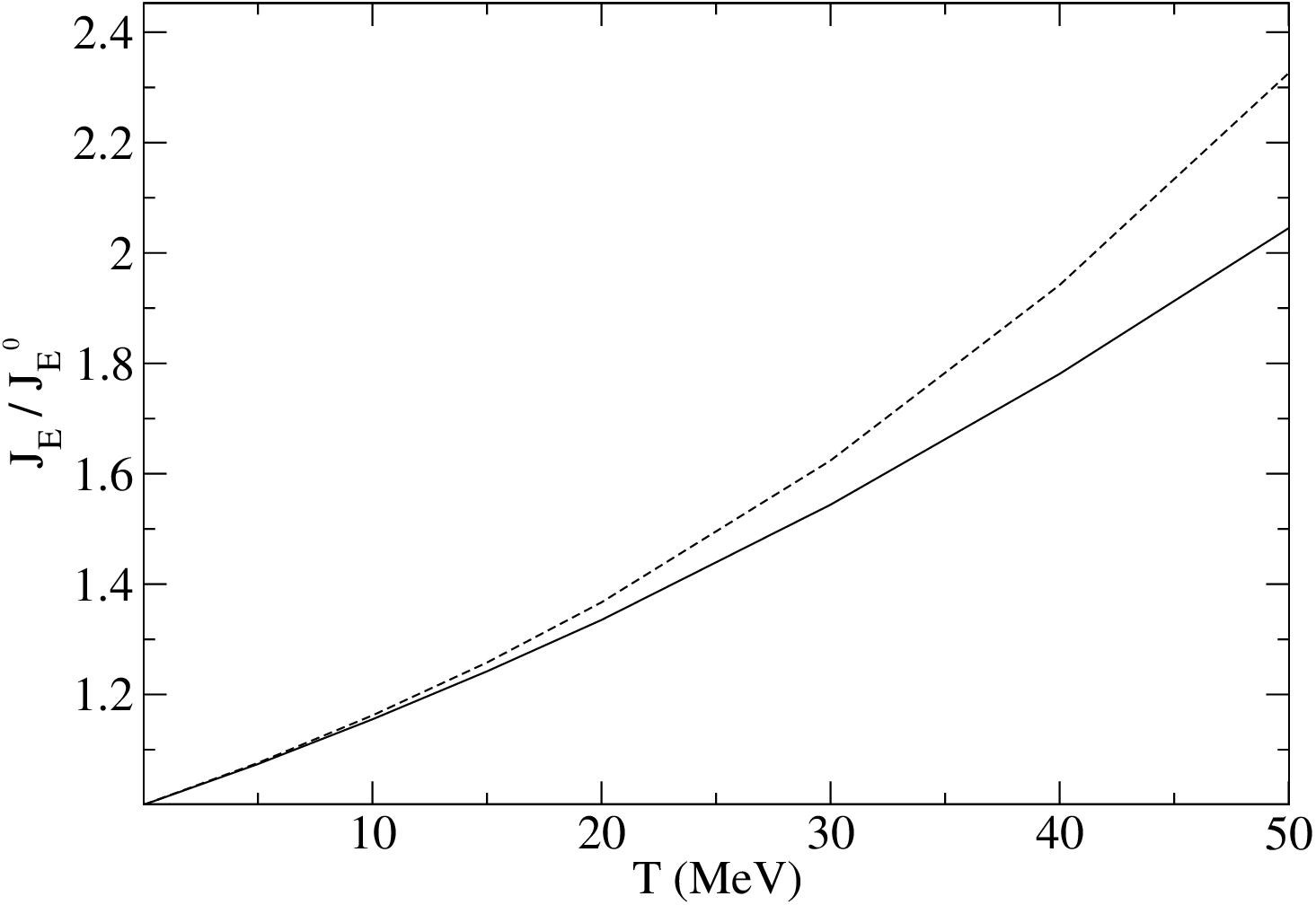}}\hss}}
\caption{Ratio of the full energy flux of $\nu_x$
and $\bar\nu_x$ neutrinos including weak magnetism 
and recoil $J_E$ to 
the energy flux without weak magnetism and recoil $J_E^0$
versus temperature. The solid line is the full
result Eq. (\ref{43}) over Eq. (\ref{44}). The dashed
line neglects the neutrino chemical potential $\mu_{\nu_x}$
in Eq. (\ref{43}) and corresponds to $\mu_{\nu_x} =a=0$. }
\label{fig4}
\end{figure}

We calculate the energy flux carried by $\nu_x$ by
multiplying the integrand in Eq. (\ref{36}) by $k$,
\begin{equation}
\vec J^E_{\nu_x}=-\int {d^3k\over (2\pi )^3} 
{\lambda (k)\over 3} \vec\nabla
{k \over 1+e^{(k-\mu_{\nu_x}) /T}} . \label{42}
\end{equation}
We use Eq. (\ref{41}) for $\mu_{\nu_x} \approx
 \mu_0=aT^2$
with $a=\delta_c \pi^2 /6M $ and add the $\bar\nu_x$
energy flux.  We assume the temperature $T$ has only a radial dependence
so that, 
\begin{eqnarray}
J_E = J^E_{\nu_x} +J^E_{\bar\nu_x} 
=- {\lambda_0k^2_0\over 6\pi^2} {dT\over dr} 
\left[ \int^\infty_0 {dk \over R^t_\nu}
\left( ak + {k^2\over T^2}\right) \left( e^{{k\over T} -aT} 
+ e^{aT-k/T}+2\right)^{-1} \right. \nonumber \\
\left. + \int^\infty_0 {dk\over R^t_{\bar\nu}} 
\left( {k^2\over T^2} - ak\right) 
\left( e^{{k\over T}+aT} + e^{-aT-k/T} +2\right)^{-1}\right] .
\label{43}  
\end{eqnarray} 
In the absence of weak magnetism and recoil corrections, 
$\mu_{\nu_x}\equiv 0$ and the energy flux becomes, 
\begin{equation}
J^0_E =- {\lambda_0k^2_0\over 6\pi^2}{dT\over dr}
2 \int^\infty_0 dk {k^2\over T^2} 
\left( e^{k/T} +e^{-k/T} +2 \right)^{-1}. \label{44}
\end{equation}
In Fig. \ref{fig4} we plot the ratio of the full energy
current with weak magnetism and recoil to the zeroth
order result $J_E/J^0_E$. {\it We see that weak magnetism
and recoil substantially enhances the energy flux.} This 
should raise the neutrino luminosity of the protoneutron star.
At a temperature of 30 MeV the enhancement is over 50\%.

Many supernova simulations combine $\nu_x$ and $\bar\nu_x$
into one effective species. In this approximation $\mu_{\nu_x}=0$.
Therefore, we evaluate $J_E$ in Eq. (\ref{43}) while
setting $a= 0$.  This would correspond to using a (geometric)
average for the correction factor, 
\begin{equation}
<R>^{-1} = {1\over 2} \left[ \left( R^t_\nu\right)^{-1} +
\left(R^t_{\bar\nu}\right)^{-1}\right], \label{45}
\end{equation}
for both $\nu_x$ and $\bar\nu_x$. We see that $J_E (a=0)/
J^0_E$ is slightly larger then $J_E/J^0_E$. This is because
the chemical potential somewhat reduces the number of $\bar \nu$
and hence their energy current. Even so, $J_E(a=0)$ is a much better approximation then $J_E^0$. 
\medskip

\subsection{Spectrum}

The above lepton number and energy fluxes are based on a diffusion
approximation.  This is valid inside the neutrino sphere where
$\lambda$ is much less than the size of the system. We now discuss
weak magnetism and recoil corrections to the emitted $\nu_x$
and $\bar\nu_x$ spectra.  We will use a simple Monte Carlo
model of Raffelt \cite{raf01} to estimate the emitted spectra. We emphasize
that this simple model should be checked against full simulations
that include weak magnetism and recoil.  Nevertheless,
we expect the model to provide a first orientation.

Raffelt \cite{raf01} discusses the formation of $\nu_x$ spectra. At 
high densities, reactions such as $ e^+ e^- \leftrightarrow
\nu_x \bar\nu_x $, $ \nu_x e\rightarrow \nu_xe$, or 
$NN\leftrightarrow NN \nu_x \bar\nu_x$ keep $\nu_x$ in chemical
and thermal equilibrium with the matter.  We define the energy
sphere as the approximate location where these reactions,
which all have small cross sections, become too slow to 
maintain thermal equilibrium.

Next the $\nu_x$ propagate through a scattering atmosphere
where neutrinos diffuse because of $N \nu_x\rightarrow N \nu_x$
which has a much larger cross section.  However the small
energy transfer in nucleon scattering is assumed to be too small
to maintain thermal equilibrium.  Finally, the neutrinos escape
from the neutrino sphere.

Note, the average spectrum for neutrinos and antineutrinos may be modified somewhat by
nucleon-nucleon bremstrahlung in the scattering atmosphere \cite{raf01}.  However, we expect 
weak magnetism in nucleon scattering to provide an estimate for the {\it difference} 
between the spectrum of neutrinos and antineutrinos. This should be explicitly checked in future 
work that includes weak magnetism for nucleon-nucleon bremstrahlung.

We model the energy sphere as a black body with temperature 
$T_{ES}\approx$ 12 MeV.  This high temperature is chosen so
that the final temperature of the emitted $\nu_x$ agrees
with detailed simulations.  Alternatively, $T_{ES}$
is the hot matter temperature well inside the neutrino
sphere at densities high enough ($\sim 10^{13} g(cm^3)$
so that the $\nu_x$ are thermalized.  We assume the same 
$T_{ES}$ for both $\nu_x$ and $\bar\nu_x$ because they are
produced in pairs.

We  next assume the $\nu_x$ propagate through a  scattering 
atmosphere of optical depth,
\begin{equation}
\tau (k) = {\tau_0k^2\over 12T_{ES}^2} R^t_\nu (k)\label{46}
\end{equation}
Here $\tau_0$ is the thermally averaged optical depth in the 
absences of recoil and weak magnetism corrections.
Raffelt assumes a Boltzmann distribution for the initial
thermalized neutrinos for which $<k^2> = 12T_{ES}^2$.
Note, the optical depth is the thickness of the scattering 
atmosphere in units of the energy dependent mean free path.
A value of $\tau_0 \sim$ 30 corresponds to one typical 
simulation \cite{raf01}.

Raffelt calculates the survival probability $S (\tau (k))$
for a $\nu_x$ of energy $k$ to make it through the atmosphere and escape.
Otherwise it is assumed to be scattered back to the energy 
sphere and absorbed. A fit to ref. \cite{raf01} Monte Carlo results is,
\begin{equation}
S(\tau) = \left[ 1+ {3\over 4} \tau \right]^{-1}
\left[ 1- {0.033 \over 
1+ 1.5 (\ell_\tau + .17)^2 + 0.5 (\ell_\tau + .32)^6 } \right],
\label{47}
\end{equation}
with $\ell_\tau = \log_{10} \tau$. Therefore the final 
emitted spectrum is,
\begin{equation}
f_{\nu_x} (k) = k^2e^{-k/T_{ES}} S \left( \tau (k) \right)
\label{48}
\end{equation}
For antineutrinos the optical depth is,
\begin{equation}
\bar\tau (k) = {\tau_0k^2\over12T^2_{ES}}
R^t_{\bar\nu}(k) , \label{49}
\end{equation}
and the $\bar\nu_x$ spectrum is,
\begin{equation}
f_{\bar\nu_x} (k) \approx N_0 k^2 e^{-k/T_{ES}} 
S \left(\bar\tau (k)\right) . \label{50}
\end{equation}
The normalization $N_0$ is chosen so that 
$\int_0^\infty dkf_{\bar\nu_x}
= \int_0^\infty dk f_{\nu_x}$.
This corresponds to steady state equilibrium and no net
change in the lepton number.  Note, $N_0$ is related to
the chemical potential $\mu_{\nu_x}$.

The $\nu_x$ and the $\bar\nu_x$ spectra are shown
in Fig. 5. Both of these curves differ from a Boltzmann distribution.  Nevertheless the average energy can 
be characterized by a spectral temperature,
\begin{equation}
T\equiv <k>/3. 
\label{51}
\end{equation}
Table II collects T values. We find that T for $\bar\nu_x$
is about 7\% larger than that for $\nu_x$ almost independent
of the choice of $\tau_0$. Thus weak magnetism insures that 
the $\bar\nu_x$ spectrum is hotter than that for $\nu_x$. 
At high energies $R^t_{\bar \nu} (k)$ is small.
Therefore $f_{\bar\nu}(k)$ becomes significantly larger than
$f_{\nu_x} (k)$.  This is shown in Fig. 5 as an increasing
ratio of $f_{\bar\nu_x} (k)$ to $f_{\nu_x} (k)$.  Thus
the high energy tail in the spectrum is expected to be antineutrino rich.

\begin{figure}
\vbox to 5in{\vss\hbox to 8in{\hss {\includegraphics{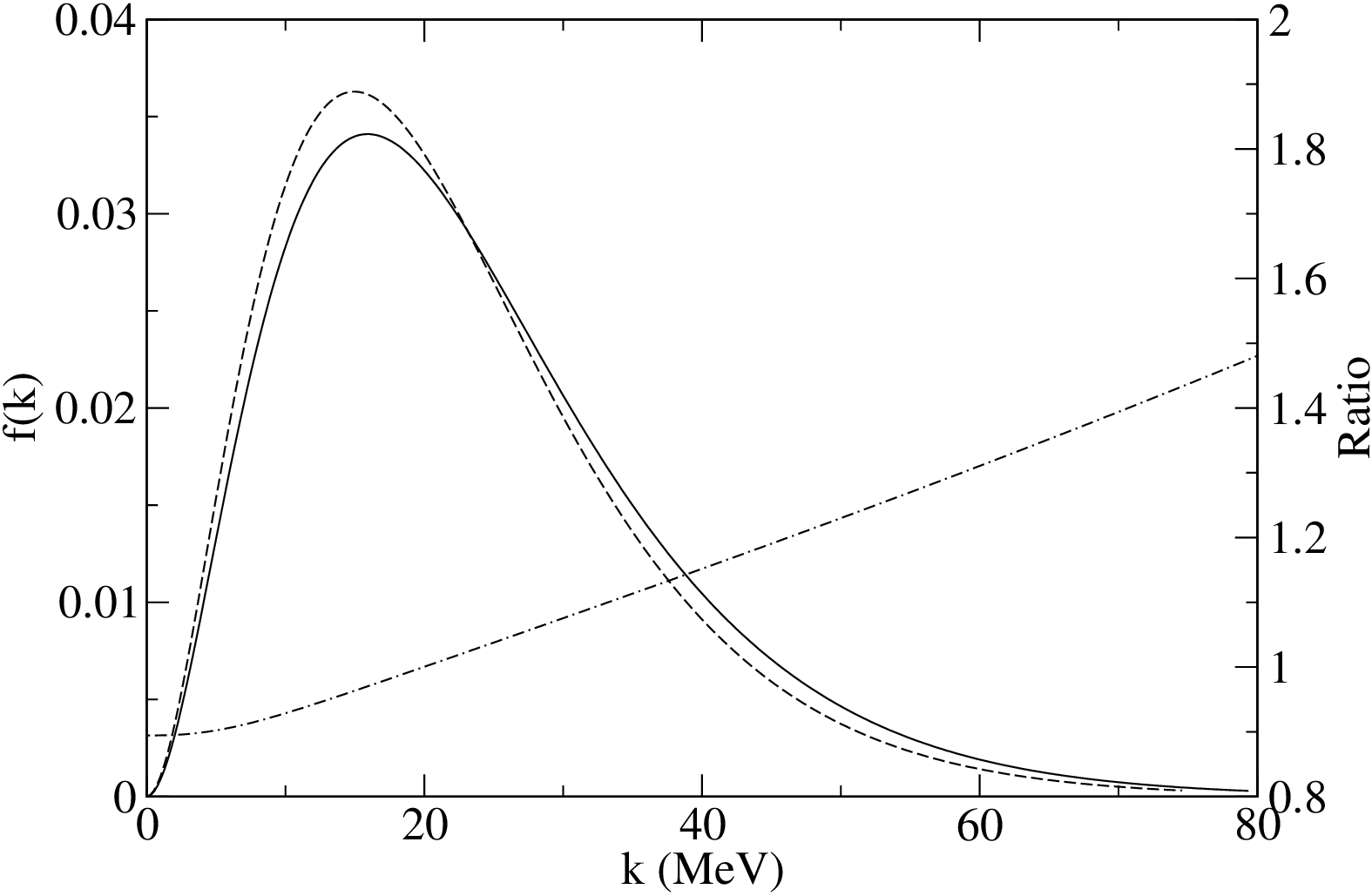}}\hss}}
\caption{Spectrum of muon or tau antineutrinos $\bar\nu_x$ 
(solid curve) and neutrinos $\nu_x$ (dashed curve) versus
neutrino energy $k$, see Eqs. (\ref{48}) and (\ref{50}).
The ratio of the $\bar\nu_x$ to $\nu_x$  spectrum is 
shown by the dot dashed curve using the right hand scale.}
\label{fig5}
\end{figure}

The difference between the $\nu_x$ and the $\bar\nu_x$ 
spectra in Fig. \ref{fig5} is a macroscopic manifestation 
of charge conjugation, C, violation in the standard model.  If the
weak interactions had conserved C then $R^t_\nu = R^t_{\bar\nu}$.
In principle, this difference is directly observable.  
However, it may be very difficult  to distinguish a detected
$\nu_x$ from a $\bar\nu_x$. Instead, a superposition 
of the $\nu_x$ and the $\bar\nu_x$ spectra may be more easily
observable. Weak magnetism, by separating the $\nu_x$ and 
the $\bar\nu_x$ spectra and by increasing the high energy tail of 
the $\bar\nu_x$ may lead to an observable broadening 
of the combined $\nu_x$ and $\bar\nu_x$ spectrum.

\medskip
\section{Electron Antineutrinos}

The reduction in $\bar\nu_e$ charged current interactions from 
weak magnetism and recoil can change the emitted $\bar\nu_e$
spectrum, decrease the electron number current and increase the 
$\bar \nu_e $ energy flux.  These changes, in turn, impact the 
$Y_e$ in the neutrino driven wind which is a possible site for 
$r-$process nucleosynthesis \cite{hor99,hor01}.  Finally these changes could
be important for analyzing the detected $\bar \nu_e$
events from SN1987A or a future galactic supernova.

\medskip
\subsection{Lepton number current}

The increased $\bar\nu_e$ mean free path from recoil and 
weak magnetism will increase the $\bar \nu_e$ number
current or decrease the total  electron lepton number current
$J_e = J_{\nu_e}-J_{\bar\nu_e}$. This can increase the $\nu_e$
density in the star. For simplicity, consider a situation where
$\mu_{\nu_e}$ is near zero in a simulation without recoil or weak
magnetism. Then in steady state equilibrium, we expect recoil and weak
magnetism to lead to a $\mu_{\nu_e}$ of order the 
$\mu_{\nu_x}$  from section III Eq. (\ref{41}).   This chemical
potential will lead to an increase in the electron
fraction because, in $\beta$  equilibrium, 
$\mu_e + \mu_n = \mu_p + \mu_{\nu_e}$.  This will lead to a small change in the electron fraction of order,
\begin{equation}
Y_e \sim Y^0_e \left[1 + {3\mu_{\nu_e} \over \mu^0_e } \right].
\label{52}
\end{equation}
Here $\mu^0_e $ is the electron chemical potential and $Y_e^0$ the electron fraction without weak magnetism and $Y_e$ is
the new electron fraction. Using Eq. (\ref{41}) for $\mu_{\nu_e}$
at T = 10 MeV, $\mu_e^0$ = 15 MeV one has,
\begin{equation}
Y_e \sim Y_e^0 (1 + .12)\label{53}
\end{equation}
an increase of 10\%. Note, Eqs. (\ref{52},\ref{53}) slightly overestimate the change in $Y_e$ because the charged current opacity for $\bar\nu_e$ increases with $Y_e$.  Nevertheless, weak magnetism could lead to a modest 
increase in $Y_e$.

In principle, weak magnetism will increase the lepton number 
diffusion time scale. However this time scale is very 
sensitive to temperature. We also expect weak magnetism to increase the 
$\bar\nu_e$ energy flux in a similar way to Fig. 4. 
With weak magnetism, the protoneutron star should cool faster
and the lower temperatures will decrease the lepton number diffusion
time.  These contrary effects should be investigated in a full 
simulation.

\medskip

\subsection{Spectrum of $\bar\nu_e$}

We consider the following simple model \cite{raf01} to estimate the 
change in the $\bar\nu_e$ spectrum with weak magnetism.  Let the
$\bar\nu_e$ neutrino sphere be at a temperature $T_0$ without 
weak magnetism. For simplicity we assume a Boltzmann spectrum
$e^{-k/T_0}$.

Now we include weak magnetism and recoil described by
$R_{\bar\nu}(k)$. This reduction in the opacity will shift
the neutrino sphere inwards to higher densities and temperatures.
We define a new energy dependent temperature $T(k)$ to include this shift,
\begin{equation}
T(k) \approx \left[ {1\over R_{\bar\nu} (k)} \right]^u T_0 .
\label{54}
\end{equation}
Here $u$ is the ratio of temperature to density gradients,
\begin{equation}
u = {d {\rm ln}T\over d {\rm ln}r}/ \left[ {d {\rm ln}\rho\over d {\rm ln}r} -1 \right],
\label{55}
\end{equation}
for the protoneutron star at the neutrino sphere.
The extra -1 in the denominator follows because the optical
depth involves a path integral from the neutrino sphere to
infinity \cite{raf01}. Realistic values of $u$ could be $\sim$ 0.25
to 0.35 \cite{raf01}. 

The energy dependent temperature $T(k)$ defines the emitted 
spectrum, 
\begin{equation}
f_{\bar\nu_e} (k) \approx k^2 e^{-k/T(k)} 
\label{56}
\end{equation}
If we expand $R_{\bar\nu} (k) \approx 1-\delta_c
k/M$, with $\delta_c = 4(c_v + F_2)c_a / (c^2_v + 3c_a^2)+3\approx 
7.22$ and expand $f_{\bar\nu_e}$ to first order in $k/M$, 
\begin{equation}
f_{\bar\nu_e}(k) \approx k^2 \left[ 1+ {u\delta_ck^2\over MT_0}\right]
e^{-k/T_0} .
\label{57} 
\end{equation}
We define a spectral temperature $T_{\bar\nu_e}$ as, 
\begin{equation}
T_{\bar\nu_e} \equiv <k>/3 \approx T_0 
\left[1+ 8u \delta_c T_0/M \right] ,
\label{58}
\end{equation}
For $u$ = 0.3 and $T_0 =$ 5 MeV one has,
\begin{equation}
T_{\bar\nu_e} \approx 1.090\, T_0.
\label{59}
\end{equation}
Thus weak magnetism and recoil can increase the emitted $\bar\nu_e$
average energy by of order 10\%. It is important to check our 
simple model with full simulations.

Figure 6 shows the full spectrum, Eq. (\ref{57}). We see that 
weak magnetism and recoil shift the strength to higher energies
and increase the high energy tail significantly. This is 
shown by a large ratio of the spectrum with weak magnetism
to $k^2e^{-k/T_0}$ at high energies.

\medskip
\subsection{Conditions outside the gain radius}
 
We now discuss some possible implications of weak magnetism
on neutrino heating, electron fraction and neutrino cooling
in the low density region outside the gain radius. The gain radius 
is the point were cooling from neutrino emission balances 
heating from neutrino 
absorption and scattering. The net amount of heating
outside the gain radius could be important for the success
of the explosion.

The absorption of $\bar\nu_e$ outside the gain radius will
be reduced by the smaller cross section from weak magnetism.
However the cross sections grows with the square of the energy
so the new absorption rate may be proportional to,
\begin{equation}
{\rm Rate} \sim \left( {T_{\bar\nu_e} \over T_0} \right)^2 
<R_{\bar\nu_e} (k) >  . 
\label{60}
\end{equation}
The first factor could increase the rate by about 20\% 
while the second factor $< R_{\bar\nu_e} > \sim $ 0.8 
could decrease the rate by a similar amount. Therefore,
the net rate of $\bar\nu_e$ absorption might be little 
changed. Thus the electron fraction, which depends on the 
absorption rate, should not change greatly.  However, the 
$\bar\nu_e$s absorbed have a higher energy so the net heating 
should increase by about 10\% because of the higher average energy in 
Eq. (\ref{59}).

This assumes the luminosity of the $\bar\nu_e$ changed 
only because of the change in $T_{\bar\nu_e}$.  
If the luminosities of $\nu_e$ and  $\bar\nu_e$ increased 
further because of the increased energy fluxes in Fig. 4
then the total heating could be even larger.

Finally we discuss cooling from positron
capture outside the gain  radius, 
\begin{equation}
e^+ + p \rightarrow n+ \bar\nu_e . 
\label{61}
\end{equation}
The rate for Eq. (\ref{61}) will be reduced by weak
magnetism,
\begin{eqnarray}
{\rm Rate} \propto \int^\infty_0 dk k^2 k^3 R_{\bar\nu_e}
(k) e^{-k/T} ,\nonumber \\
\approx \int^\infty_0 dk k^5 \left( 1 - \delta_c {k\over m}\right)
e^{-k/T} , \nonumber \\
= 5 ! T^6 \left( 1- {6\delta_c T\over M} \right) . 
\label{62}
\end{eqnarray}
At $T$ = 2 MeV the factor in parenthesis reduces the cooling rate by 9\%. 
Thus weak magnetism can change the $\bar\nu_e$ luminosity
to increase the heating while, at the same time, reducing the 
cross section to decrease the cooling.  These results should 
be checked with full simulations.

\medskip
\section{Conclusion}

In this paper we examine recoil and weak magnetism
corrections to $\nu$-nucleon interactions. 
These are important because they are present at 
all densities, even at the relatively low densities
near the neutrino sphere.  Furthermore, the 
corrections are model independent. We calculate simple, 
exact correction factors Eqs. (\ref{12},\ref{22},\ref{32})
to include recoil and weak magnetism in supernova simulations.

Perhaps the most important effect of weak magnetism and recoil is to 
increase energy fluxes of both $\bar\nu_x$
and  $\bar\nu_e$ antineutrinos.  We calculate, in a 
diffusion approximation, an increase of order 15\%
in the total energy flux for temperatures near 10 MeV. 
This should raise the neutrino luminosity.

Weak magnetism and recoil will also change the emitted spectrum of 
$ \bar\nu_x$ and $\bar\nu_e$. We estimate that $\bar\nu_x$ will be 
emitted about 7\% hotter than $\nu_x$ because $\bar\nu_x$ have longer 
mean free paths.  Likewise weak magnetism may increase the $\bar\nu_e$
temperature by of order 10\%.  This increase in temperature
coupled with the increase in neutrino luminosity should overcome the 
reduced absorption cross section and increase
the heating in the low density region outside of the neutrino sphere.
This, in turn, could be important for the success of an explosion.

We find large corrections.  However, supernova simulations are very 
complicated with many degrees of feedback.  Therefore, it is important 
to check our results with a full simulation that includes Boltzmann
neutrino transport and weak magnetism corrections.

\acknowledgements
We thank John Beacom, Hans-Thomas Janka and Georg Raffelt for extensive discussions of the physics.  We thank the Institute for Nuclear Theory in Seattle for its hospitality when this work was started and acknowledge financial support from DOE grant DE-FG02-87ER40365.

\vskip 1in

\begin{tabular}{llll} \hline
\multicolumn{4}{c}{Table I Coupling Constants}   \\
Reaction    & $c_v$   & $c_a$  & $F_2$ \\ \hline
$ \nu p \rightarrow \nu p$ 
& $ {1 \over 2} -2 \sin^2 \theta_{w} \approx$ 0.035 
& $g_a /2 \approx$  0.630  
& $ {1\over 2} ( \mu_p -\mu_n ) - 2 \sin^{2} \theta_{w} \mu_p \approx
  $ 1.019 \\
  $\nu n \rightarrow \nu n$ 
& $-{1\over 2}$
& $-g_a /2 \approx$  -0.630 
& $ - {1\over 2} ( \mu_p -\mu_n ) - 2 \sin^{2} \theta_{w} \mu_n \approx
  $ -.963 \\
$ \left. { \nu_e n\rightarrow e^-p \atop \bar\nu_e p 
\rightarrow e^+ n } \right\}$      
& 1 
& $g_a \approx$ 1.260 
& $\mu_p -\mu_n \approx $ 3.706 \\
\label{table1}
\end{tabular}

{Here $g_a \approx$ 1.260, $\sin^2\theta_w \approx$
0.2325, $\mu_p=$ 1.793 and $\mu_n=-$ 1.913. } 
\vskip 1in

\begin{tabular}{lll} \hline
\multicolumn{3}{c}{Table II Corrections to $\nu$ Opacities}   \\
 &  &  Correction  \\ \hline
1. &&  Phase space \\
2. &   &  Matrix element \\
   &a. &  recoil	\\
   &b. &  weak magnetism \\	
   &c. &  form factors \\
   &d. &  strange quarks \\
3. &   &  Pauli blocking \\ 
4. &   &  Fermi/thermal motion of initial nucleons \\
5. &   &  Coulomb interactions \\ \hline
6. &   &  Mean field effects \\
7. & & NN Correlations in RPA \\
8. & & NN Correlations beyond RPA \\
9. & &  Meson exchange currents \\
10. & &  Other components such as hyperons \\
11. & &  Other phases such as meson condensates or quark matter \\
12. & &  Corrections from superfluid/ superconductor pairing \\
13. & &  Nonuniform matter  \\ \hline 
14. & &   Magnetic field effects \\
\label{table2}
\end{tabular}
\vskip 1in

\begin{tabular}{llll} \hline 
\multicolumn{4}{c}{Table III Spectral temperature} 
$T=<k>/3$   \\
$\tau_0$ & $T_{\nu_x}$ (MeV)  & $T_{\bar\nu_x}$  (MeV) & $T_{\bar\nu_x}/T_{\nu_x}$\\
\hline
3 &8.75 & 9.24 & 1.056 \\
10  &7.48 & 7.96 & 1.064\\
30  &6.49 & 6.92&  1.066 \\
100 &5.64 &6.04 & 1.071 \\
300 &5.09 & 5.46&1.073\\
\label{table3}
\end{tabular}
\vfill\eject 


\vfill\eject

\appendix
\section{Low Density Limit of Burrows and Sawyer}

In this appendix we discuss the low density limit of Burrows and Sawyer's calculations \cite{bur98} and show that it reduces to our phase space result, $d\sigma_{ps}/d\Omega$, in Eq. (\ref{6}).  Burrows and Sawyer start by considering a pure vector interaction.  Their Eq. (2) for the differential rate of neutrino scattering ${\bf k} \rightarrow {\bf k^\prime}$ is,
\begin{equation}
{d^2\Gamma\over d\omega d{\rm cos}\theta}={G^2\over 4\pi^2} {k^\prime}^2 [1-f_\nu(k^\prime)]\Lambda^{00}S(q,\omega).
\label{A1.1}
\end{equation}
Here the momentum transfered to the neutrons is ${\bf q}={\bf k}-{\bf k^\prime}$ and the energy transfered is $\omega=k-k^\prime$.  The neutrino distribution is $f_\nu(k)$ and $\Lambda^{00}$ is the neutrino tensor, Eq. (3) of ref. \cite{bur98}.  The dynamic structure function $S(q,\omega)$ for the nuetrons is,
\begin{equation}
S(q,\omega)=2\int{d^3p\over (2\pi)^3} f(p) [1-f(|{\bf p}+{\bf q}|)] 2\pi \delta(\omega+\epsilon_p - \epsilon_{p+q}),
\label{A1.2}
\end{equation}
where $f(p)$ is the neutron distribution function and $\epsilon_p$ the energy of a neutron of momentum $p$.

One can recover our Eq. (\ref{6}) by, (a) neglecting the final state neutrino pauli blocking, (b) integrating Eq. (\ref{A1.1}) over $\omega$, (c) assuming that the density is low enough so that $1-f({\bf p}+{\bf q})\approx 1$ in Eq. (\ref{A1.2}),
\begin{equation}
\int_0^\infty d\omega S(q,\omega)\approx 2\int {d^3p\over (2\pi)^3} f(p) 2\pi,
\nonumber
\end{equation} 
\begin{equation}
=2\pi \rho_n.
\label{A1.3}
\end{equation}
Here $\rho_n$ is the neutron density.  Finally (d) we assume, in the limit of low density, that the $\omega$ depdendence of $S(q,\omega)$ is sharply peaked near, $\omega = k-k^\prime$, with $k^\prime$ given by our Eq. (\ref{2}).  Adding a simillar expression for the axial interactions gives,
\begin{equation}
\int_0^\infty {d\Gamma\over d\omega d{\rm cos}\theta} d\omega \rightarrow 2\pi \rho_n {d\sigma_{ps}\over d\Omega},
\label{A1.4}
\end{equation}
with $d\sigma_{ps}/d\Omega$ given by our Eq. (\ref{6}).

\medskip
\section{Single Nucleon Form Factors}

In this appendix we collect simple 
parameterizations of the single nucleon form factors for use
in  Eq. (\ref{13}). We write the weak form factors in terms of the 
electromagnetic Dirac, $F^i_1$, and Pauli, $F_2^i$, form factors for 
$i=p$ or $n$.

\medskip

\leftline{1) Reaction $\nu p \rightarrow \nu p$}

\begin{equation}
c_v = \left( {1\over 2}  - 2 \sin^2 \theta_w\right) F_1^{(p)}
-{1\over 2} F_1^{(n)}
\label{A2.1}
\end{equation}
\begin{equation}
c_a= {g_a\over 2} \left( 1 + 3.53 \tau \right)^{-2} 
\label{A2.2}
\end{equation}

\begin{equation}
F_2 = \left( {1\over 2} - 2\sin^2 \theta_w \right) 
F_2^{(p)} - {1\over 2} F_2^{(n)}
\label{A2.3}
\end{equation}

\medskip
\leftline{2) Reaction to $\nu n \rightarrow \nu n$}

\begin{equation}
c_v = \left( {1\over 2} - 2 \sin^2 \theta_w\right) F^{(n)}_1
- {1\over 2} F_1^{(p)} 
\label{A2.4}
\end{equation}
\begin{equation}
c_a = - {g_a\over 2} \left( 1 + 3.53 \tau\right)^{-2}
\label{A2.5}
\end{equation}
\begin{equation}
F_2 = \left( {1\over 2} - 2\sin^2\theta_w\right) F_2^{(n)} 
- {1\over 2} F_2^{(p)}
\label{A2.6}
\end{equation}

\medskip

\leftline{Reaction $\nu_e n\rightarrow e^- p$}

\begin{equation}
c_v = F_1^{(p)} - F_1^{(n)} 
\label{A2.7}
\end{equation}
\begin{equation}
c_a = g_a \left(1+3.53 \tau \right)^{-2} 
\label{A2.8}
\end{equation}
\begin{equation}
F_2 = F_2^{(p)} - F_2^{(n)}
\label{A2.9}
\end{equation}

Here, see Eq. (\ref{14}), 
\begin{equation}
\tau = Q^2 / 4M^2 , 
\label{A2.10}
\end{equation}
\begin{equation}
F_1^{(p)} = \left[ 1 + \tau \left( 1+\lambda_p \right) \right]
G / \left( 1 + \tau \right) , 
\label{A2.11}
\end{equation}
\begin{equation}
F^{(p)}_2 = \lambda_p G/\left( 1 + \tau \right), 
\label{A2.12}
\end{equation}
\begin{equation}
F_1^{(n)} \approx \tau \lambda_n \left( 1- \eta \right)
G / \left(1 + \tau \right)
\label{A2.13}
\end{equation}
\begin{equation}
F^{(n)}_2 = \lambda_n \left( 1 + \tau\eta \right) 
G/ \left( 1 +\tau\right) , 
\label{A2.14}
\end{equation}
with 
\begin{equation}
\lambda_p = 1.793, \lambda_n =- 1.913,
\label{A2.15}
\end{equation}
\begin{equation}
\eta  = \left( 1 + 5.6 \tau \right)^{-1}, 
\label{A2.16}
\end{equation}
and finally, 
\begin{equation}
G = \left( 1+ 4.97 \tau \right)^{-2} 
\label{A2.17}
\end{equation}
These parameterizations are from Ref. \cite{mus92}.
\clearpage


\begin{figure}
\vbox to 5in{\vss\hbox to 8in{\hss {\includegraphics{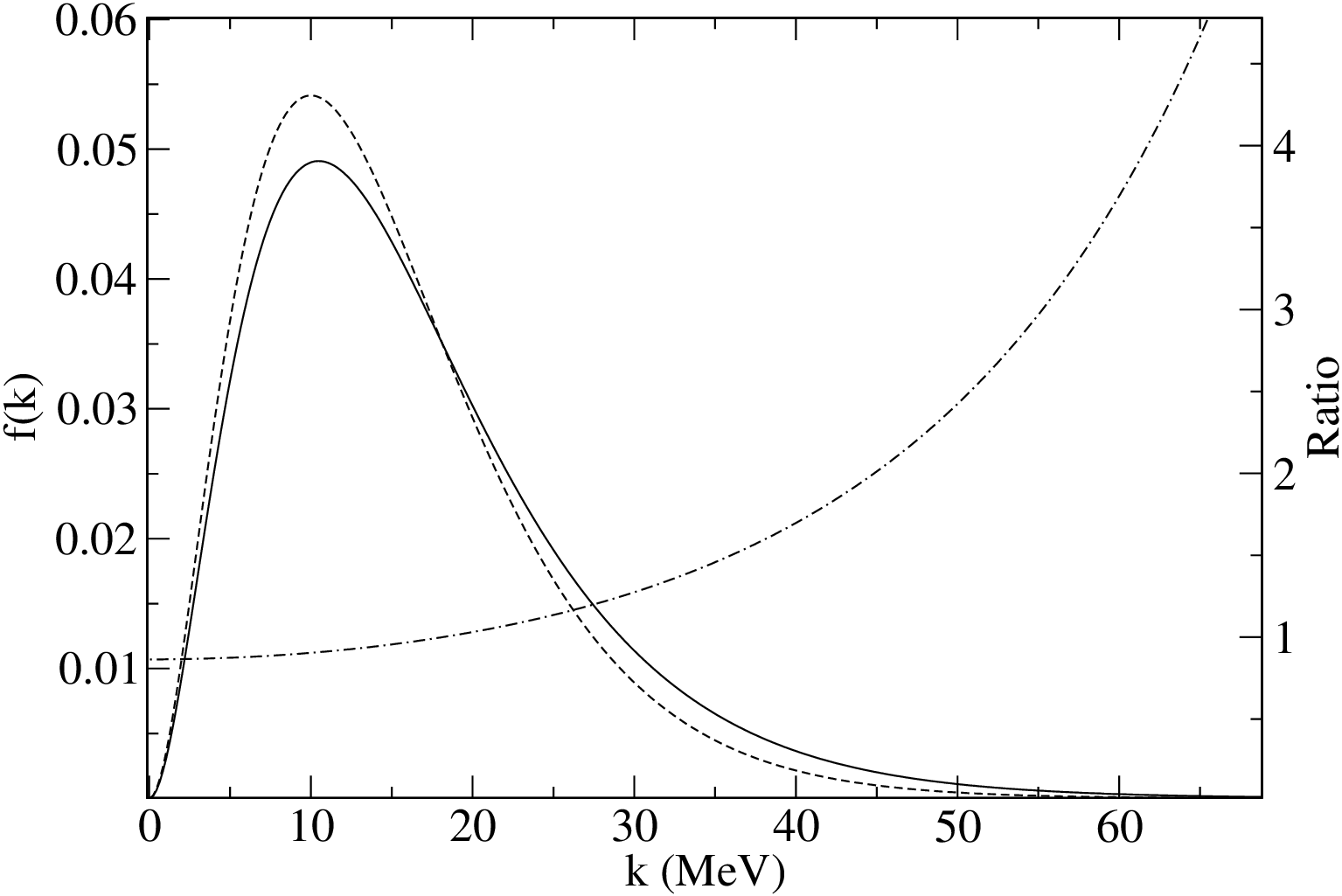}}\hss}}
\caption{Spectrum of electron antineutrinos $\bar\nu_e$ 
versus neutrino energy $k$. The solid curve includes weak
magnetism and recoil, Eq. (\protect{\ref{56}}), while these are 
neglected in the dashed curve.  The ratio of solid 
to dashed curves is plotted as the dot-dashed curve using 
the right hand scale.}
\label{fig6}
\end{figure}

\end{document}